\documentclass{aa}

\usepackage{graphicx}
\usepackage{txfonts}

\begin{document}

\title{Diverse lifestyles of bar-like galaxies and their coevolution with the brightest galaxy in the most massive
cluster of TNG50}
\titlerunning{Diverse lifestyles of bar-like galaxies in the most massive cluster of TNG50}

\author{Ewa L. {\L}okas
}

\institute{Nicolaus Copernicus Astronomical Center, Polish Academy of Sciences,
Bartycka 18, 00-716 Warsaw, Poland\\
\email{lokas@camk.edu.pl}}

\abstract{
Clusters can provide propitious environments for bar formation in galaxies. This work studies the formation and
evolution of 15 bar-like galaxies in the most massive cluster of the TNG50 simulation from the IllustrisTNG suite. The
selection includes galaxies from the last simulation output from well-resolved subhalos with a strongly prolate stellar
component. Eleven galaxies form or strongly enhance their bars during a pericenter passage around one or more
progenitors of the brightest cluster galaxy (BCG). Two form their bars early as a result of minor mergers, one via an
interaction with another massive galaxy, and one via disk instability. The bar formation times differ considerably,
ranging between 3-11 Gyr. The lengths of the bars also differ, ranging between 2-6 kpc, and do not correlate with the
amount of tidal forcing experienced. All galaxies have at least one pericenter passage around a BCG progenitor, but the
number of interactions varies strongly and is reflected in the different amount of mass stripping the galaxies
experience. Most bar formation events take place before the BCG is fully formed. In three cases, they occur just before
different progenitors of the BCG merge. For six bar-like galaxies, the merger events leading to the final formation of
the BCG cause significant changes of their orbits. Their diverse evolutionary histories illustrate the different paths
to bar formation in clusters and emphasize the complex nature of the process, which includes coevolution with BCG
progenitors.}


\maketitle

\section{Introduction}

\nolinenumbers

Bars in galaxies are generally believed to form as a result of the inherent instability of their disks \citep{Hohl1971,
Ostriker1973, Athanassoula2003}. Recently, however, another possible scenario for their formation has received
considerable attention, namely the one involving interactions with other objects. Since interactions occur more
frequently at the early stages of galactic evolution, this scenario has been evoked especially in the context of
recent discoveries of high-redshift bars with James Webb Space Telescope (JWST) \citep{Guo2023, Costantin2023,
Amvrosiadis2025, Lokas2025a, Lokas2025c} and measurements of nonzero bar fractions at early epochs \citep{LeConte2024,
Guo2025, Geron2025}.

Interactions leading to bar formation in disky galaxies can take many forms and can involve dwarf galaxies orbiting a
Milky Way-like host \citep{Lokas2014, Lokas2015, Gajda2017, Gajda2018}, a flyby of an object of similar mass
\citep{Noguchi1987, Gerin1990, Miwa1998, Berentzen2004, Lokas2018}, or a Milky Way-like galaxy orbiting a cluster
\citep{Mastropietro2005, Lokas2016}. In all these configurations, studied mostly through controlled $N$-body
simulations, the mechanism of bar formation is the same and related to the tidal force acting on the disk, which
distorts it and induces more radial orbits of the stars. The effect is larger for more massive hosts, smaller
pericenter distances, and more prograde orientations of the disk with respect to the orbit. Sufficient tidal force
leads to a fast transformation of almost the whole stellar component from an oblate to a prolate shape. The resulting
shape of the object is bar-like rather than barred, since little of a disk remains in which the bar could be embedded.

The formation of tidal bars in the cluster environment is particularly promising because of the strong tidal force
exerted by the brightest cluster galaxy (BCG). Since most cluster members are expected to experience at least one
pericenter passage around their BCG, this environment should efficiently produce tidal bars, and we expect an increased
fraction of barred galaxies toward the cluster center. Currently, there is no clear observational evidence supporting
this relationship, although early studies appeared to support this hypothesis \citep{Thompson1981, Andersen1996,
Barazza2009}. However, later studies found little or no significant dependence of the bar fraction on the
clustercentric distance \citep{Mendez2010, Lansbury2014, Cervantes2015, Lokas2016} and pointed to other factors as
decisive in bar formation and properties, such as the stellar mass, size and morphological type of the galaxy
\citep{Tawfeek2022, Aguerri2023}.

Simulations of galaxy formation and evolution can help clarify the situation. In \citet{Lokas2016}, we considered a
very idealized configuration in which an $N$-body model of a Milky Way-like galaxy was placed on different orbits in a
Virgo-like cluster and evolved for 10 Gyr. The Milky Way galaxy was composed of a disk and a dark matter halo, whereas
the Virgo cluster was approximated as a single, preexisting, although live, dark halo. These controlled experiments
demonstrate that, in such an environment, bars form much faster than in isolation, particularly on tighter orbits.

Cosmological simulations of galaxy formation, such as IllustrisTNG \citep{Springel2018, Marinacci2018, Naiman2018,
Nelson2018, Pillepich2018}, provide an ideal tool to place this scenario in a more realistic context. These simulations
follow the evolution of both dark matter and the baryonic component in boxes of different size and resolution (TNG50,
TNG100, and TNG300) by solving for gravity and magnetohydrodynamics and are supplemented by subgrid prescriptions for
processes such as star formation, stellar feedback, and black hole feedback. In such simulations, the bars are formed
alongside other phenomena, such as galaxy clustering into groups and BCG formation via mergers of massive galaxies,
which complicates the picture compared to the evolution of a single galaxy in a preexisting cluster. In
\citet{Lokas2020}, I studied the tidal evolution of galaxies in the most massive cluster of TNG100 and showed that bars
can indeed form via interactions with a BCG, presenting several examples.

Interesting candidates for tidally induced bars occur among bar-like galaxies. These are not bars embedded in disks, as
previously studied using Illustris and IllustrisTNG \citep{Peschken2019, Rosas2020, Rosas2022, Rosas2025, Zhou2020,
Zhao2020}, but rather objects whose overall stellar component can be approximated as a prolate spheroid. In
\citet{Lokas2021}, I studied a sample of bar-like galaxies in the TNG100 simulation, selected among well-resolved
objects (total stellar mass above $10^{10}$ M$_\odot$), with a single condition that the intermediate-to-longest axis
ratio $b/a$ of the stellar component (within two stellar half-mass radii, $2 r_{1/2}$) be lower than 0.6. The sample
contained 277 bar-like galaxies, which could be divided into three classes based on the origin of the bar and the
subsequent evolution. In class A galaxies (77 objects, 28\% of the sample), an interaction with a larger perturber,
typically a central galaxy of a group or cluster, induced the bar. In classes B and C (27\% and 45\% of the sample,
respectively), a minor merger, a small satellite, or disk instability induced the bars. While class B galaxies were
partially stripped of mass, those of class C evolved in isolation, retaining their mass content.

\citet{Lokas2025b} provides a detailed study of the 77 bar-like galaxies of class A whose strong bars were induced by
an interaction with more massive objects in a cluster. For the entire subsample, the time of bar formation strongly
correlated with and was typically slightly greater than the time of the pericenter passage around the host. All
galaxies were strongly stripped of dark matter and gas, and their rotation was diminished. Larger pericenter distances
typically required higher host masses to transform the galaxies. Despite difficulties in interpreting some cases
involving mergers and multiple interactions, the results confirmed, in the cosmological context, the reality of tidal
bar formation in cluster environments previously studied using controlled simulations and generalized the previous
study \citep{Lokas2020}, which considered only the most massive cluster of TNG100.

This work uses a higher-resolution simulation of the IllustrisTNG project, namely TNG50 \citep{Nelson2019b,
Pillepich2019}, to study the properties of bar-like galaxies in the most massive cluster of this simulation box. The
higher resolution allows the selection of objects with total stellar mass above $10^{9}$ M$_\odot$ (corresponding to
around $10^4$ stellar particles), which remain suitable for morphological analysis. This choice enables the
construction of a sample of 15 bar-like galaxies in a single cluster, whereas only a few such objects per cluster could
be identified and followed in TNG100. The evolutionary paths of these galaxies can then be studied in the same
environment, allowing a meaningful comparison between their histories.

The paper is organized as follows. Section~2 presents the main properties of the 15 selected bar-like galaxies and
their evolution, focusing on the different measures of shape and the mass content. Section~3 describes the individual
evolutionary paths of the galaxies, emphasizing their differences. Section~4 discusses the classification of the
diverse evolutionary histories into groups with common features and demonstrates the coevolution of the bar-like
galaxies and the forming cluster, in particular the build-up of its BCG through a series of mergers.

\begin{table*}
\caption{Properties of the bar-like galaxies at $z=0$}.
\label{properties}
\centering
\begin{tabular}{c c c c c c c c c c c c c}
\hline\hline
ID  & $2 r_{1/2}$ & $M_*$ & $M_{\rm dm}$ & $b/a$ & $c/a$ & $T$ & $A_2$ & $f$ & $t_{\rm bf}$ & $A_{\rm 2,max}$ & $R_{\rm
bar}$ & $d_{\rm ID0}$ \\
    &  [kpc]  & [$10^{10}$M$_\odot$] & [$10^{11}$M$_\odot$]        &   &    &       & & & [Gyr] &   & [kpc]  & [kpc]\\
\hline
13   &    2.8 &   2.12	 & 2.84   & 0.53  & 0.42  & 0.88  & 0.47 &  0.27 & 11.3  & 0.61   & 3.8   & 1793  \\
14   &    4.7 &   2.76	 & 1.51   & 0.59  & 0.41  & 0.79  & 0.45 &  0.28 &  8.1  & 0.58   & 4.9   & 1291  \\
15   &    2.7 &   3.33	 & 0.81   & 0.57  & 0.40  & 0.81  & 0.30 &  0.23 & 10.2  & 0.65   & 4.4   & 701  \\
23   &	  2.0 &   1.82	 & 1.09   & 0.56  & 0.48  & 0.90  & 0.37 &  0.10 &  5.9  & 0.55   & 2.3   & 428  \\
31   &	  2.2 &   1.49	 & 0.44   & 0.56  & 0.48  & 0.88  & 0.40 &  0.10 &  7.6  & 0.60   & 2.1   & 443  \\
34   &	  3.2 &   1.25	 & 0.55   & 0.56  & 0.43  & 0.84  & 0.38 &  0.12 &  6.5  & 0.53   & 6.0   & 720  \\
35   &	  3.2 &   1.43	 & 0.34   & 0.49  & 0.41  & 0.91  & 0.48 &  0.10 &  3.6  & 0.64   & 4.1   & 503  \\
38   &	  3.0 &   1.39	 & 0.19   & 0.60  & 0.46  & 0.82  & 0.43 &  0.12 &  5.0  & 0.55   & 3.5   & 488  \\
42   &	  1.6 &   1.39	 & 0.07   & 0.55  & 0.47  & 0.89  & 0.38 &  0.07 &  3.1  & 0.58   & 2.1   & 133  \\
55   &	  3.5 &   0.87	 & 0.21   & 0.53  & 0.43  & 0.89  & 0.49 &  0.05 &  8.6  & 0.59   & 5.1   & 158  \\
64   &	  3.1 &   0.67	 & 0.20   & 0.55  & 0.47  & 0.89  & 0.44 &  0.08 &  9.2  & 0.55   & 4.2   & 313  \\
71   &	  2.7 &   0.67	 & 0.09   & 0.53  & 0.41  & 0.87  & 0.46 &  0.06 &  9.7  & 0.57   & 4.5   & 261  \\
72   &	  2.1 &   0.68	 & 0.07   & 0.56  & 0.46  & 0.87  & 0.41 &  0.04 &  8.4  & 0.55   & 2.9   & 114  \\
87   &	  3.1 &   0.52	 & 0.08   & 0.56  & 0.45  & 0.86  & 0.39 &  0.04 &  6.7  & 0.51   & 5.1   & 148  \\
308  &	  0.9 &   0.10	 & 0.003  & 0.50  & 0.50  & 0.99  & 0.44 &  0.06 &  3.1  & 0.62   & 3.3   & 596  \\
\hline
Median &  2.8 &   1.39   & 0.21   & 0.56  & 0.45  & 0.88  & 0.43 &  0.10 &  7.6  & 0.58   & 4.1   & 443  \\
\hline
\end{tabular}
\end{table*}

\section{Formation and properties of the bar-like galaxies}

This work uses the TNG50 simulation \citep{Nelson2019b, Pillepich2019}, the highest-resolution simulation of the
IllustrisTNG project, performed in an approximately 50 Mpc box. The simulation includes baryonic particles with masses
of $8.5 \times 10^4$ M$_\odot$ and dark matter particles with masses of $4.5 \times 10^5$ M$_\odot$, with corresponding
softening scales of 0.074 and 0.288 kpc, respectively. The simulation data comprise 100 outputs, which provide
sufficient time resolution to follow galaxy evolution. The data are publicly available, well documented, and easily
retrievable, as described in \citet{Nelson2019a}. The subhalos corresponding to galaxies are found using the Subfind
algorithm \citep{Springel2001}, which assigns mass to them by locating locally overdense, self-bound particle groups
within a larger parent group. Galaxies are identified by their subhalo ID numbers, which differ at subsequent outputs.
Here, galaxies are referred to by their IDs in the last simulation output, corresponding to the present time ($t =
13.8$ Gyr, $z = 0$).

The most massive cluster of galaxies formed in this simulation has a BCG corresponding to the subhalo ID0. The total
mass of this object at the end of the simulation is $2 \times 10^{14}$ M$_\odot$, which includes the stellar mass of
$M_* = 5.4 \times 10^{12}$ M$_\odot$, the gas mass of $M_{\rm gas} = 2.3 \times 10^{13}$ M$_\odot$, and the dark mass
of $M_{\rm dm} = 1.7 \times 10^{14}$ M$_\odot$. The BCG formed through major mergers of the cluster's  most massive
galaxies. This formation process significantly affected the evolution of other galaxies, including the bar-like
galaxies discussed here.

\begin{figure*}
\centering
\includegraphics[width=15cm]{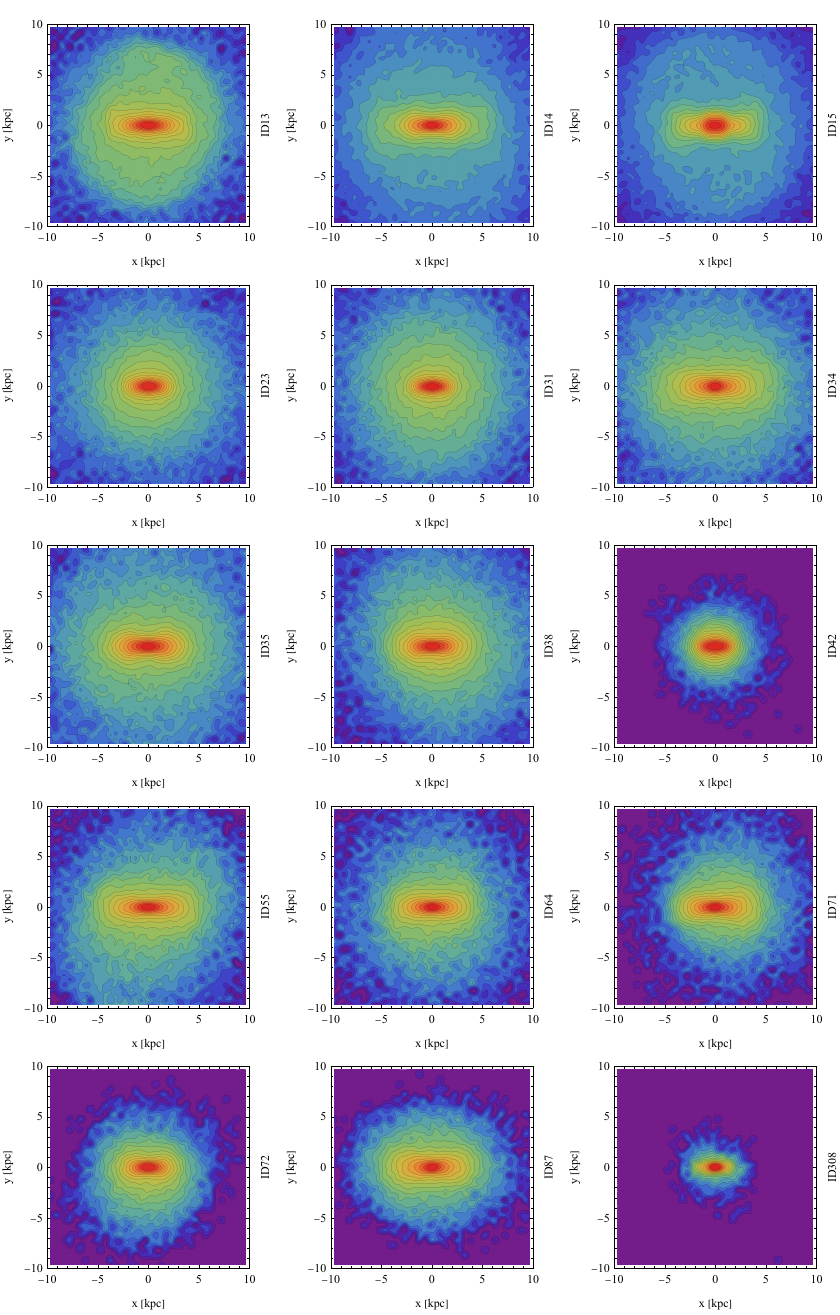}
\caption{Surface density maps of the stellar component of 15 bar-like galaxies in face-on view at $z = 0$.
Galaxies are identified by their subhalo ID numbers at $z = 0$, shown on the right of each panel. The color scale is
adjusted to the maximum and minimum density of each galaxy and the contours are equally spaced in log surface density.}
\label{surden}
\end{figure*}

The bar-like galaxies used in this study were selected among cluster members with a total stellar mass above $10^{9}$
M$_\odot$ at the final simulation output. This value corresponds to about $10^4$ stellar particles at TNG50 resolution,
which enables morphological analysis. The TNG50 catalogs list 178 objects meeting this mass criterion, which reduces to
133 galaxies after removing dark-matter-free artifacts. To select bar-like galaxies, I applied a single criterion: the
intermediate-to-longest axis ratio $b/a$ of the stellar component within two stellar half-mass radii, $2 r_{1/2}$, had
to be lower than 0.6. This condition was satisfied by 15 galaxies, whose IDs are listed in the first column of
Table~\ref{properties}. The second column lists the values of their stellar half-mass radii, $r_{1/2}$.
Fig.~\ref{surden} shows their images in terms of the surface density of the stars in the face-on view at the present
time.

The axis ratio calculations followed the method of \citet{Genel2015} and agree with the values estimated by the
Illustris team, provided in the Supplementary Data Catalogs of stellar circularities, angular momenta, and axis ratios.
The eigenvalues of the stellar mass tensor provided the axis values after aligning each galaxy with its principal axes
and calculating three components ($i$=1,2,3): $M_i = (\Sigma_j m_j r^2_{j,i}/\Sigma_j m_j)^{1/2}$, where $j$ enumerates
over stellar particles, $r_{j,i}$ is the distance of stellar particle $j$ in the $i$-axis from the center of the
galaxy, and $m_j$ is its mass. Sorting the eigenvalues so that $M_1 < M_2 < M_3,$ yielded the shortest-to-longest axis
ratio $c/a = M_1/M_3$ and the intermediate-to-longest axis ratio $b/a = M_2/M_3$. A useful combination of the axis
ratios is the triaxiality parameter $T = [1-(b/a)^2]/[1-(c/a)^2]$, which measures the shape with a single value. Values
of  $T < 1/3$ correspond to strongly oblate objects, $T > 2/3$ to strongly prolate objects, and intermediate values
correspond to triaxial shapes.

\begin{figure*}
\centering
\includegraphics[width=18cm]{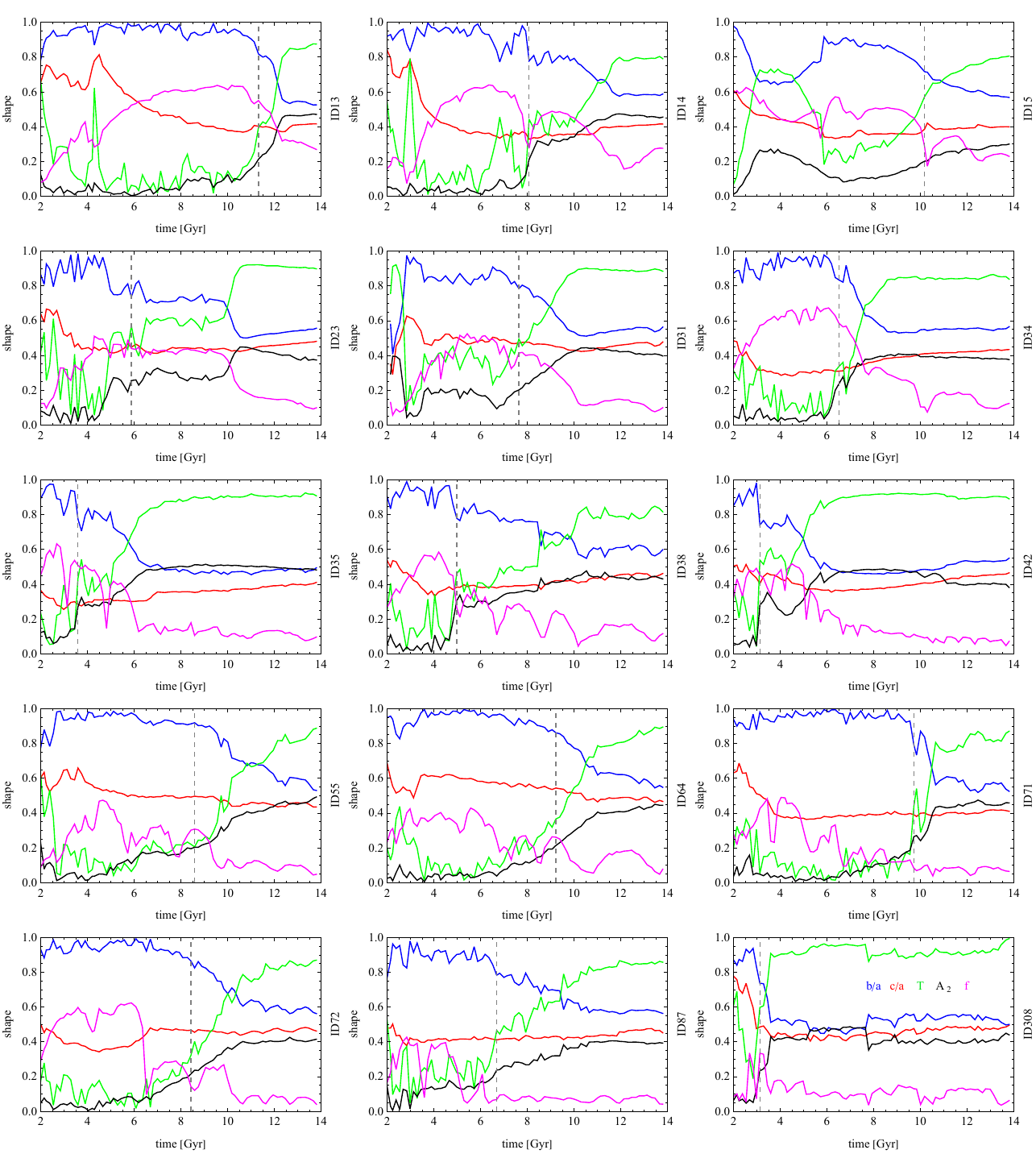}
\caption{Evolution of different measures of shape over time for 15 bar-like galaxies. Lines show the axis ratios
$b/a$ (blue), $c/a$ (red), the triaxiality parameter $T$ (green), the rotation parameter $f$ (magenta), and the bar mode
$A_2$ (black). Galaxies are identified by their subhalo ID numbers at $z = 0$, shown on the right of each panel.
Vertical dashed lines indicate the time of bar formation.}
\label{shape}
\end{figure*}

In addition to shape measures, it is useful to consider the kinematics of the stellar component, particularly the
amount of rotational support in the galaxy. This was measured for TNG50 galaxies using the rotation parameter, $f$,
defined as the fractional mass of all stars with the circularity parameter $\epsilon > 0.7$, where $\epsilon=J_z/J(E)$,
$J_z$ is the specific angular momentum of the star along the angular momentum of the galaxy, and $J(E)$ is the maximum
angular momentum of stellar particles \citep{Genel2015}. This parameter reliably measures the amount of rotation in the
galaxy, with $f > 0.4$ considered characteristic of disks \citep{Joshi2020}. Figure~\ref{shape} shows the measures of
shape and kinematics as a function of time for all 15 bar-like galaxies, and the parameter values at the end of the
evolution are listed in Table~\ref{properties} (columns 5-7, 9).

The evolution of the bar is best quantified using the commonly adopted measure of bar strength
\citep{Athanassoula2002}, namely the $m=2$ Fourier mode of surface density projected along the short axis. This is
given by $A_2 (R) = | \Sigma_j m_j \exp(2 i \theta_j) |/\Sigma_j m_j$, where $\theta_j$ is the azimuthal angle of the
$j$th star, $m_j$ is its mass, and the sum goes up to the number of particles in a given radial bin. Figure~\ref{shape}
shows the measurements of the bar mode using stars within two stellar half-mass radii, $2 r_{1/2}$ (black line). The
bar formation time $t_{\rm bf}$ corresponds to the moment when $A_2 (< 2 r_{1/2})$ crosses 0.2 and remains above this
value until the end of the evolution; a vertical dashed line marks it in each case. Table~\ref{properties} lists the
values of $A_2 (< 2 r_{1/2})$ at the end of the evolution and the bar formation times in the eighth and tenth columns,
respectively.

The axis ratio $b/a$ and the bar mode $A_2$ provide alternative measures of shape: $A_2$ uses a projected distribution
of stars, whereas $b/a$ takes into account the 3D shape (because it implicitly assumes $c<b$). The criterion $b/a <
0.6$ is convenient initially, as $b/a$ values are provided in the TNG50 data release. While $b/a$ and $A_2$ remain
strongly correlated at all times, with $A_2=0.2$ corresponding roughly to $b/a=0.8$, the $A_2$ threshold is preferable
to characterize the bars at early times because it appears more stable.

The evolution of shape and kinematic measures in Fig.~\ref{shape} exhibits common features across all galaxies.
Although the bar formation times differ, bar formation always produces a strongly increasing value of triaxiality $T$
and a decreasing value of $b/a$, while rotation decreases due to the more radial orbits of stars within the bar.
Additionally, the evolution of the rotation parameter $f$ often shows sudden drops, which sometimes coincide with bar
formation. These decreases in rotation typically result from tidal forces acting on the galaxy during interactions with
a massive companion. These interactions also produce sudden drops in the galaxy's bound mass.

\begin{figure*}
\centering
\includegraphics[width=18cm]{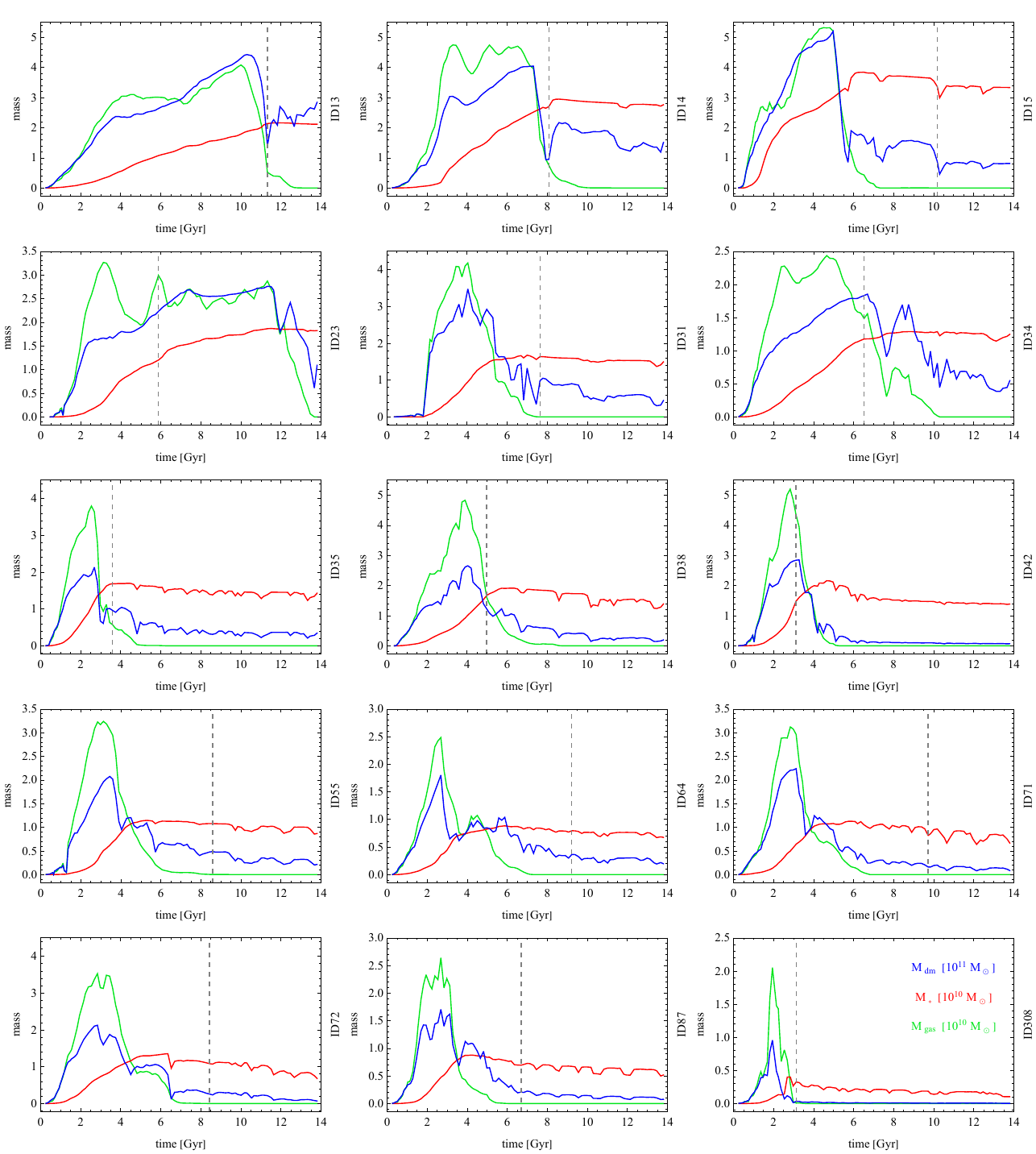}
\caption{Evolution of the total mass over time for 15 bar-like galaxies. Lines show the total galaxy mass in dark
matter (blue), stars (red), and gas (green). The dark masses are given in units of $10^{11}$ M$_\odot$, while
the stellar and gas masses are in units of $10^{10}$ M$_\odot$. Galaxies are identified by their subhalo ID numbers
at $z = 0$, shown on the right of each panel. Vertical dashed lines indicate the time of bar formation.}
\label{mass}
\end{figure*}

To verify this, Fig.~\ref{mass} shows the evolution of each galaxy's  total mass in three components: dark matter,
stars, and gas. All galaxies grew in mass until experiencing a sudden drop, particularly in the dark component and the
gas, while the stellar mass was much less affected. Stellar masses reached a maximum and ceased growing due to the lack
of gas, then declined slowly. The sudden drops in dark mass content coincide with the drops in the rotation parameters
shown in Fig.~\ref{shape}. The total stellar and dark masses of the galaxies at the end of the evolution are listed in
the third and fourth columns of Table~\ref{properties}. The galaxies lost their gas by that time, except for ID13 and
ID15, which retained a small gas mass of $10^{7}$ M$_\odot$.

\begin{figure*}
\centering
\includegraphics[width=18cm]{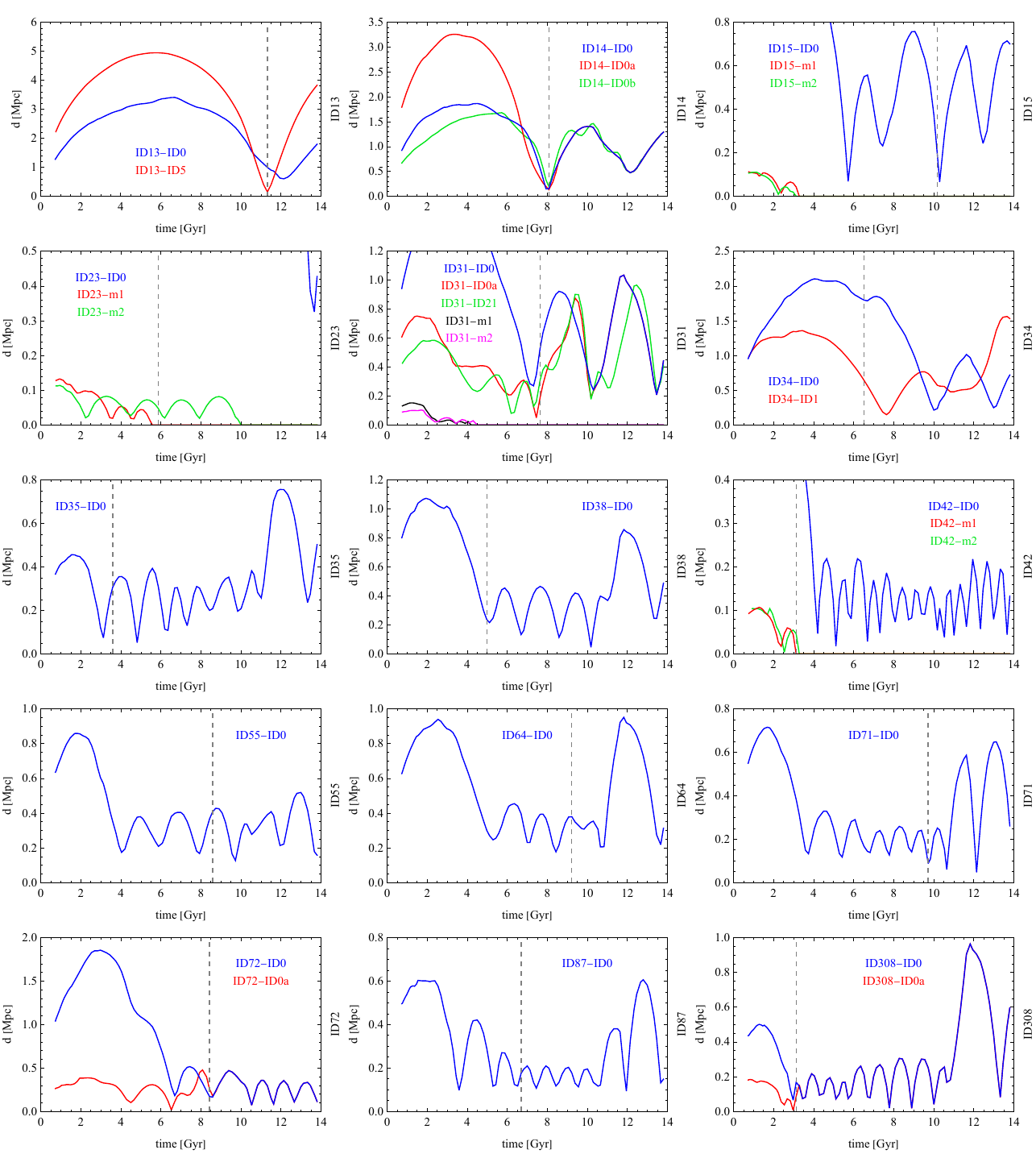}
\caption{Distances from perturbers as a function of time for 15 bar-like galaxies. Different colored lines show
the distances from distinct perturbers, with data for ID0 shown in blue in all panels. Galaxies are
identified by their subhalo ID numbers at $z = 0$,  shown on the right of each panel.
Vertical dashed lines indicate the time of bar formation.}
\label{neighbors}
\end{figure*}

To identify the sources of the tidal force, I searched the simulation data for neighbors of the 15 galaxies at times
when their shape and mass content evolved most strongly. This allowed me to identify perturbers that were massive and
close enough to affect the galaxies under study. Figure~\ref{neighbors} shows the relative distances between each
galaxy and the perturbers most likely to influence its evolution. The minima of the distances correspond to the
pericenter passages, where the interactions are strongest. In all panels, the blue line shows the distance to the most
massive progenitor of the BCG (ID0) and the figure demonstrates that each galaxy experienced at least one pericenter
passage around it. The present-day distances of the galaxies from ID0 are listed in the last column of
Table~\ref{properties}. However, this interaction was not the most important one for some galaxies and, in a few cases,
other galaxies and mergers played an important role. Details of these interactions are described in the next section.

\begin{figure}
\centering
\includegraphics[width=8.9cm]{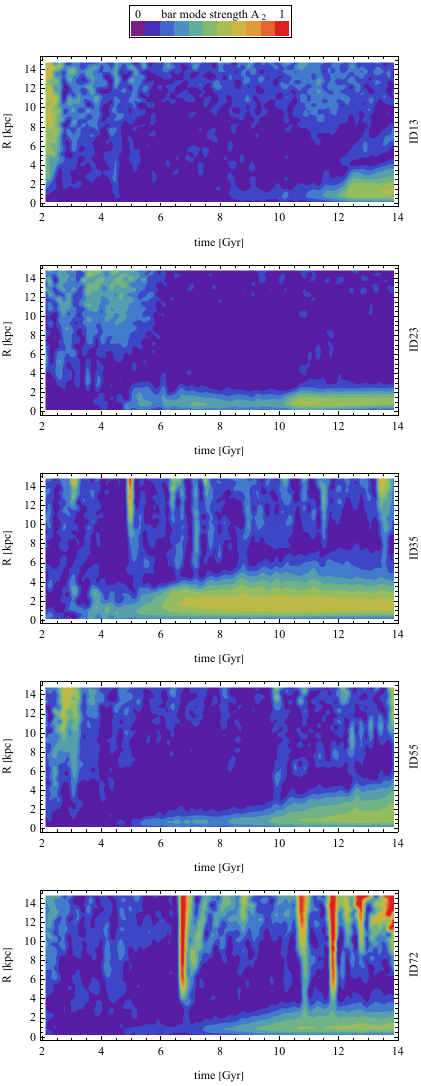}
\caption{Evolution of the profile of the bar mode, $A_2 (R)$, over time for five galaxies from the first columns of
Figs.~\ref{surden}-\ref{neighbors}. The galaxies are identified by their subhalo ID numbers
at $z = 0$ given on the right of each panel. The reddest color corresponds to regions with no stars bound to the galaxy.}
\label{a2modes1}
\end{figure}

\begin{figure}
\centering
\includegraphics[width=8.9cm]{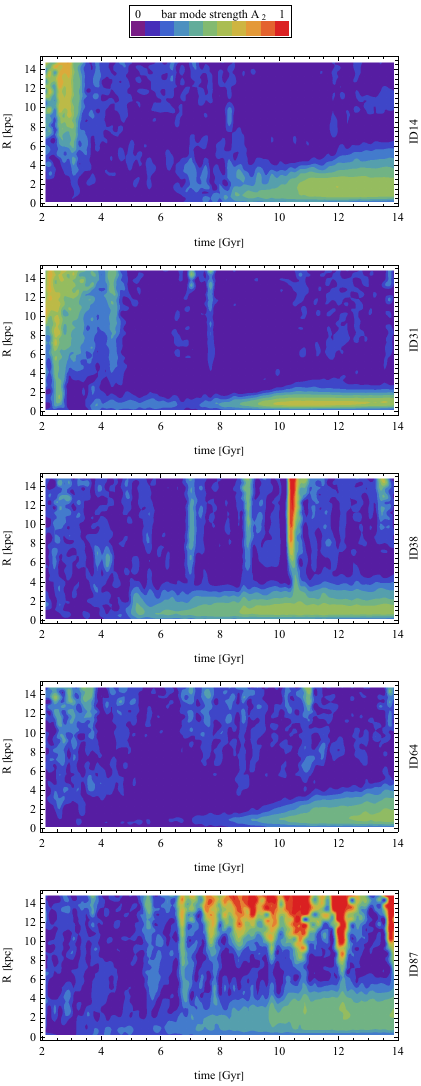}
\caption{Same as Fig.~\ref{a2modes1}, but for five galaxies from the second columns of
Figs.~\ref{surden}-\ref{neighbors}.}
\label{a2modes2}
\end{figure}

\begin{figure}
\centering
\includegraphics[width=8.9cm]{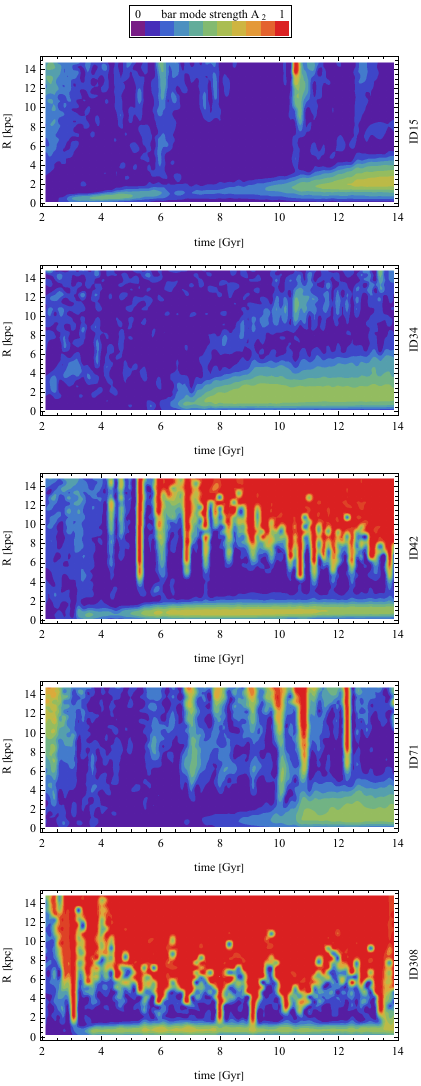}
\caption{Same as Fig.~\ref{a2modes1}, but for five galaxies from the third columns of
Figs.~\ref{surden}-\ref{neighbors}.}
\label{a2modes3}
\end{figure}

Finally, Figs.~\ref{a2modes1}-\ref{a2modes3} illustrate the detailed evolution of galaxy shapes through the bar mode
profiles as a function of time, $A_2 (R, t)$. The bar mode was measured in bins of $\Delta R = 0.5$ kpc out to 15 kpc
in cylindrical radius. The color-coded maps show a persistent bar-like shape near the galaxy center and the elongations
of each galaxy at larger radii caused by interactions. Redder colors indicate stronger elongations, which reach their
extremes at the outskirts where tidal tails appear. These outer regions become increasingly sparsely populated, and the
reddest shade corresponds to regions where no stars are bound to the galaxy due to tidal stripping. These measurements
can also be used to estimate the bar length. The typical shape of the $A_2 (R)$ profile for a barred galaxy rises to a
maximum value, $A_{\rm 2, max}$, and then decreases, except during temporary strong tidal distortions, when a secondary
rise occurs at larger radii. The bar length can be estimated as the radius $R_{\rm bar}$ where $A_2 (R)$ drops to
$A_{\rm 2, max}/2$. Table~\ref{properties} lists the values of $A_{\rm 2, max}$ and $R_{\rm bar}$ at the end of the
evolution in the 11th and 12th columns.

\section{Individual evolutionary tracks}

This section presents the different evolutionary histories of the galaxies inferred from the results presented in
Figs.~\ref{surden}-\ref{a2modes3}. The galaxies are described in the same order, in which they are listed in
Table~\ref{properties}.

\paragraph{ID13} The galaxy was a late and short-time visitor to the cluster and is now at its outskirts. It approached
ID0 and its companions around 11 Gyr and its bar formed at 11.3 Gyr during the interaction with ID5 at a pericenter
distance of $r_{\rm peri} = 150$ kpc; ID5 had a mass of $3.6 \times 10^{11}$ M$_\odot$ at that time. The tidal force
was not particularly strong, but it was likely sufficient to strengthen the small bar-like perturbation already present
at 8.5 Gyr. The bar was further enhanced at 12.1 Gyr when the galaxy interacted with ID0 (with $r_{\rm peri} = 602$
kpc), which had a mass of $2 \times 10^{14}$ M$_\odot$ at that time.

\paragraph{ID14} The bar was induced at 8.1 Gyr after a strong interaction ($r_{\rm peri} = 151$ kpc) with the most
massive progenitor of ID0, which had a mass of $8.6 \times 10^{13}$ M$_\odot$ at that time. Simultaneously, the galaxy
interacted with two smaller progenitors of ID0 (ID0a and ID0b in Fig.~\ref{neighbors}), which later merged with the
BCG. The masses of these progenitors were both below $10^{12}$ M$_\odot$, while their pericenters were comparable to
the one of the main perturber.

\paragraph{ID15} This galaxy formed a small bar as early as 3 Gyr, likely due to minor mergers (m1 and m2 in
Fig.~\ref{neighbors}, both with masses above $10^{9}$ M$_\odot$ before the interactions) and flybys of small
satellites. The first pericenter passage around ID0 (with $r_{\rm peri} = 69$ kpc) at 6 Gyr destroyed the bar. A third
pericenter around the same perturber at 10 Gyr (with $r_{\rm peri} = 66$ kpc) strengthened the bar again, allowing it
to survive until today. The satellites of ID15 that may have contributed to the initial bar formation later merged with
ID0 during and after the first pericenter.

\paragraph{ID23} The bar in this galaxy began to form around 5 Gyr and was strongly enhanced at 10 Gyr, likely due to
mergers with small satellites (m1 and m2 in Fig.~\ref{neighbors}). The galaxy only recently entered the cluster and
experienced one pericenter passage around ID0 ($r_{\rm peri} = 325$ kpc), which was sufficient to remove all its gas
and most of its dark matter but left the bar unaffected.

\paragraph{ID31} This galaxy formed a small bar as early as 3.5 Gyr, likely as a result of interactions and mergers
with smaller satellites. An interaction with the more massive galaxy ID21 at 6.5 Gyr weakened the bar. The permanent
bar formed shortly afterward, at 7.6 Gyr, following interactions with two massive progenitors of ID0.

\paragraph{ID34} Inherent disk instability or very small mergers triggered the formation of the bar in this galaxy at
6.5 Gyr. Interaction with the very massive galaxy ID1 at 7.6 Gyr increased the bar's strength. The galaxy entered the
cluster much later and experienced two pericenter passages around ID0, at 10 and 13 Gyr, which had little effect on the
bar.

\paragraph{ID35} The bar formed at 3.6 Gyr after the first pericenter (with $r_{\rm peri} = 74$ kpc)
around the most massive progenitor of ID0 and was enhanced at the second pericenter.  The bar remained stable in its
strength and length afterward as the galaxy's orbit became more extended. This case represents a classical example of a
bar tidally induced by interaction with a cluster BCG.

\paragraph{ID38} The galaxy formed a bar at 5 Gyr during the first pericenter passage (with $r_{\rm peri} = 216$ kpc)
around the most massive progenitor of ID0 and remained on a tight orbit until this progenitor merged with another
massive galaxy, which changed the orbital configuration.  The bar strengthened during the third pericenter passage at
8.6 Gyr, while the tightest pericenter around 10 Gyr had little effect. This case represents a relatively simple
example of bar formation in the cluster environment.

\paragraph{ID42} The bar was formed very early, at 3.1 Gyr, likely due to the last minor mergers the galaxy
experienced. The bar weakened slightly after the first pericenter passage around a progenitor of ID0 and strengthened
again after the second pericenter. The galaxy remained on a very tight orbit around ID0 until the end of the evolution
and was heavily stripped of mass. The bar remained short enough to avoid strong effects from close encounters, except
for minor weakening around 11 Gyr after several particularly tight pericenter passages.

\paragraph{ID55} The galaxy started to form a bar around 6 Gyr after the second pericenter passage around a progenitor
of ID0. The bar reached the bar formation threshold at 8.6 Gyr, after the third pericenter, and strengthened further
following the fourth pericenter passage at 10 Gyr. It continued to grow steadily until the present.

\paragraph{ID64} The bar formed at 9.2 Gyr after the third pericenter passage around the most massive progenitor of
ID0 and grew steadily before and after that time, with the length slightly increasing after the next pericenter
passage at around 11 Gyr. During the formation period, the galaxy also interacted with a few other progenitors of ID0,
which makes the case less clear.

\paragraph{ID71} Bar growth began at 7.5 Gyr following the third pericenter passage around ID0. The bar fully formed at
the sixth pericenter at 9.7 Gyr and strengthened further at the next, very tight pericenter at 10.7 Gyr. Another tight
pericenter at 12.1 Gyr did not strongly affect its properties.

\paragraph{ID72} A small elongation was already present in this galaxy at 5 Gyr, which grew after a strong interaction
with the two most massive progenitors of ID0 around 6.5-6.7 Gyr. The less massive progenitor (ID0a) exerted a stronger
tidal force because the pericenter distance in this case was much smaller ($r_{\rm peri} = 18$ kpc). The bar grew
steadily and fully formed at 8.4 Gyr during the second pericenter around the more massive progenitor of ID0.
Interestingly, this is also when the two progenitors of ID0 merge. Further strong interactions with the merged ID0 did
not significantly affect the bar.

\paragraph{ID87} The bar began to form around 4 Gyr after the first pericenter passage around one of the massive
progenitors of ID0 and fully formed at 6.7 Gyr after the third pericenter. The galaxy remained on a tight orbit around
ID0 and was heavily stripped, while subsequent pericenter passages made the bar longer and stronger.

\paragraph{ID308} This is the smallest and most heavily stripped galaxy in the sample, although its dark matter mass
reached almost $10^{11}$ M$_\odot$ before interactions. The bar formed very early, at 3.1 Gyr, through interactions
with a massive progenitor of ID0 and a less massive one (ID0a), which soon merged with the more massive progenitor. The
tidal force of the less massive progenitor, with a mass of $5 \times 10^{11}$ M$_\odot$ at 3 Gyr was greater because it
passed ID308 at a very small $r_{\rm peri} = 6$ kpc, whereas the more massive progenitor, with a mass $1.5 \times
10^{13}$ M$_\odot$ had $r_{\rm peri} = 65$ kpc. The galaxy remained on a very tight orbit around ID0 and the subsequent
pericenter passages slightly strengthened the bar around 5 Gyr and weakened it at 8 Gyr, but it survived until the
present.

\section{Discussion}

This study examined the properties of bar-like galaxies in the most massive cluster of TNG50. The last row of
Table~\ref{properties} lists the median values of their main parameters at the end of the evolution. In addition to
their defining feature, a strongly prolate stellar component, the only properties these galaxies share are strong mass
loss and the lack of gas and star formation. Although clusters generally contain galaxies with these features, the
bar-like sample appears to be rather extreme in this respect. In particular, only two out of 15 galaxies (13\%) in the
sample have some (very low) gas content of around $10^7$ M$_\odot$, whereas 29\% of all 133 galaxies in the cluster
have nonzero gas content, and 14\% have significant gas mass above $10^9$ M$_\odot$. None of the 15 bar-like galaxies
form stars at $z=0$, whereas 17\% of the total sample still have nonzero star formation rates. All 15 bar-like galaxies
are red with $g-r > 0.74$ mag, while 86\% of the total sample are red with $g-r>0.6$, and only 60\% are as red as the
bar-like galaxies ($g-r>0.74$). All the bar-like galaxies are thus red and dead and would be classified as typical
ellipticals unless viewed exactly face-on, in which case their shapes would appear more elongated.

The low gas content and early quenching may have facilitated bar formation. Most of the bars studied here form when the
galaxy is completely stripped of gas, which strongly suggests that the bars form more easily when the gas fraction is
low. Figure~\ref{mass} shows that, for 11 of the 15 galaxies, there is no gas left when the bar forms or its mass is
much smaller than the mass of stars. In cases when the total gas mass equals or exceeds the stellar mass at that time
(IDs: 23, 34, 38, and 42), the gas mass within $2 r_{1/2}$ (where the bar forms) remains much lower than the stellar
mass.

All 15 galaxies (except ID13) lost more than half, and often much more, of their maximum dark mass. These mass loss
features are expected because the galaxies were selected as cluster members and therefore must have interacted with the
BCG or other galaxies during their evolution. All galaxies experienced at least one, and often multiple, pericenter
passages around the BCG (ID0), so none are infalling into the cluster for the first time.

However, the similarities between galaxies end here. Beyond these definition-related properties, the galaxies follow
very different evolutionary histories. Some galaxies experienced strong interactions with the BCG and were heavily
stripped, even in the stellar component, as shown in Figs.~\ref{surden}, \ref{mass}, and \ref{a2modes1}-\ref{a2modes3}.
This subsample includes ID42, ID71, ID72, ID87, and ID308. Among these, ID308 experienced particularly strong
stripping and is the smallest and least massive galaxy in the sample. In this respect, it is similar to the
Sagittarius-like object studied in \citet{Lokas2024}, except that it evolved around a BCG rather than a Milky Way-like
galaxy. The remaining galaxies of the sample evolved on less tight orbits. The ID13 and ID23 galaxies exemplify the
other extreme: the least affected galaxies, each with only a single pericenter passage around ID0 in their history. The
galaxies span a wide range of distances from the cluster center, from 114 kpc (ID72) to 1.8 Mpc (ID13). Strongly
stripped galaxies typically lie closer (the last column of Table~\ref{properties}), though this is not always the case,
as discussed below.

The galaxies also exhibit strong differences in bar formation times, bar properties, and formation mechanisms. The bar
formation times (tenth column of Table~\ref{properties}) range from 3.1 to 11.3 Gyr, with the median value of 7.6 Gyr.
In the cases of bars induced by interactions, the bars form between 0-0.8 Gyr after pericenter passage, with an
uncertainty of 0.1-0.2 Gyr following from the time difference between the available simulation outputs. The ID42 and
ID308 galaxies both formed their bars at the earliest time of 3.1 Gyr, but through different mechanisms: minor mergers
in ID42 and interaction with a progenitor of ID0  in ID308. These formation times are later, and the bar lengths are
shorter than those of the best candidates for the analogs of high-redshift bars discovered with JWST selected from
TNG100  in \citet{Lokas2025a, Lokas2025c}.

All bars are strong at the final time, with $A_{\rm 2, max} > 0.5$ and a median of 0.58, reflecting the selection
criterion of a strongly prolate shape, $b/a < 0.6$. However, the bars  have very different lengths, from 2.1 to 6 kpc
(median 4.1 kpc), and show little correlation between the bar length and orbit tightness or the amount of stripping the
galaxy experienced. This is because interactions can shorten bars through tidal stripping or lengthen them when tidally
induced. The bar of ID308 provides a good example of the latter case, as it remained short for most of its history and
only elongated after the last pericenter passage. The bar pattern speeds also vary significantly and range between
$9-27$ km s$^{-1}$ kpc$^{-1}$. These values were estimated at $z=0$ using the \citet{Tremaine1984} method for the
first eight galaxies in Table~\ref{properties}. The remaining objects are strongly perturbed by recent interactions and
their pattern speeds cannot be reliably measured.

Overall, interaction with one of the progenitors of ID0 tidally induced or strongly enhanced the bars in 11 of the 15
galaxies, but only ID35 formed its bar directly after the first pericenter passage around a single most massive
progenitor of ID0. In two cases (ID23 and ID42) the formation of the bar could be related to mergers with smaller
satellites, while a passage near another massive galaxy likely drove bar formation in ID13. In ID34, inherent disk
instability likely formed the bar, although the role of minor mergers cannot be completely excluded. In ID15, the bar
formed early by mergers was destroyed at the first pericenter passage around ID0 and formed again at the third
pericenter.

Despite the different origins, the subsequent evolution of the bars was to some extent affected by the following
pericenter passages around the BCG. The only exception to this rule is ID23, which had only one very recent and distant
passage and is the only galaxy in the sample that can be described as having evolved in isolation, as evidenced by no
tidal features at large radii visible in Fig.~\ref{a2modes1}. However, even for galaxies that interacted strongly with
the progenitors of ID0, the impact of these interactions varied significantly between objects and even subsequent
pericenters for the same galaxy. This is probably due to the orientation of the bar at the pericenter, which results
from a random combination of the bar pattern speed and the galaxy's velocity on the orbit. As described in detail in
\citet{Lokas2014},  the tidal force from the perturber can weaken or strengthen the bar depending on this orientation.

The resolution of the TNG50 simulation may influence the results of this study. The selection required galaxies to have
a stellar mass above $10^9$ M$_\odot$ at the end of evolution, corresponding to about $10^4$ stellar particles, to
enable morphological analysis. Observations of cluster galaxy populations \citep{Mendez2023} indicate that barred
galaxies rarely occur below this stellar mass. This suggests that the sample considered here is representative of the
bar-like population in clusters. \citet{Frankel2022} demonstrated that the resolution mostly affects the bar pattern
speeds (which are underestimated in low-resolution simulations), rather than other properties such as the bar length.
Since the pattern speeds are difficult to measure for the bar-like objects studied in this work (as stated above), this
issue should not be of concern here. Other bar properties, particularly their evolutionary histories, which are the
focus of this work, should not be very sensitive to resolution.

In summary, this analysis reveals that the formation and evolution of bar-like galaxies in the cluster environment
differ strongly from the idealized scenario of controlled simulations, such as presented in \citet{Lokas2016}, where a
Milky Way-like galaxy was placed on an orbit around a preexisting cluster. The bar-like galaxies in this study
witnessed the formation of the BCG and coevolved with its progenitors. Most bar formation events and subsequent
interactions occurred before the BCG was fully formed. The bar-like galaxies interacted with multiple progenitors of
ID0, including two (ID31, ID72,and ID308) or three at once (ID14). In three cases (ID14, ID72, and ID308), these
interactions and bar formation occurred immediately before the progenitors of ID0 merged. For six bar-like galaxies,
the merger events leading to the final formation of the BCG later on (around 11 Gyr) led to significant alterations of
their orbits around the BCG. These galaxies were ejected on more extended orbits which led to fewer (if not less tight)
pericenter passages. This may explain why some of the bar-like galaxies (in particular ID308) survived until the
present.

The significance of this complicated scenario and its implications for the efficiency of bar formation in clusters
deserve further study. It remains unclear what would happen if the configuration were simpler, that is, if the BCG
formed earlier and the other galaxies interacted with a single object rather than several of its massive progenitors.
Studying clusters with BCGs that have different formation histories could address this question, although drawing firm
conclusions may remain challenging due to random variations in cluster properties such as mass and composition.

\begin{acknowledgements}
I am grateful to the IllustrisTNG team for making their simulations publicly available and to the anonymous
reviewer for useful comments. This work was supported in part by the National Science Centre of Poland with grant
2025/57/B/ST9/00321. Computations for this work were performed using the computer cluster at the Nicolaus Copernicus
Astronomical Center of the Polish Academy of Sciences (CAMK PAN).
\end{acknowledgements}


\begin{thebibliography}{}

\bibitem[{Aguerri et al.}(2023)]{Aguerri2023} Aguerri, J. A. L., Cuomo, V., Rojas-Roncero, A., Morelli, L. 2023,
	A\&A, 679, A5
\bibitem[{Amvrosiadis et al.}(2025)]{Amvrosiadis2025} Amvrosiadis, A., Lange, S., Nightingale, J. W., et al.
	2025, MNRAS, 537, 1163
\bibitem[{Andersen}(1996)]{Andersen1996} Andersen, V. 1996, AJ, 111, 1805
\bibitem[{Athanassoula}(2003)]{Athanassoula2003} Athanassoula, E. 2003, MNRAS, 341, 1179
\bibitem[{Athanassoula \& Misiriotis}(2002)]{Athanassoula2002} Athanassoula, E., \& Misiriotis, A. 2002,
        MNRAS, 330, 35
\bibitem[{Barazza et al.}(2009)]{Barazza2009} Barazza, F. D., Jablonka, P., Desai, V., et al. 2009, A\&A, 497, 713
\bibitem[{Berentzen et al.}(2004)]{Berentzen2004} Berentzen, I., Athanassoula, E., Heller, C. H., \& Fricke, K. J.
        2004, MNRAS, 347, 220
\bibitem[{Cervantes Sodi et al.}(2015)]{Cervantes2015} Cervantes Sodi, B., Li, C., \& Park, C. 2015, ApJ, 807, 111
\bibitem[{Costantin et al.}(2023)]{Costantin2023} Costantin, L., P\'{e}rez-Gonz\'{a}lez, P. G., Guo, Y., et al.
	2023, Nature, 623, 499
\bibitem[{Frankel et al.}(2022)]{Frankel2022} Frankel, N., Pillepich, A., Rix, H.-W., et al. 2022, ApJ, 940, 61
\bibitem[{Gajda et al.}(2017)]{Gajda2017} Gajda, G., {\L}okas, E. L., \& Athanassoula, E. 2017, ApJ, 842, 56
\bibitem[{Gajda et al.}(2018)]{Gajda2018} Gajda, G., {\L}okas, E. L., \& Athanassoula, E. 2018, ApJ, 868, 100
\bibitem[{Genel et al.}(2015)]{Genel2015} Genel, S., Fall, S. M., Hernquist, L., et al. 2015, ApJ, 804, L40
\bibitem[{Gerin et al.}(1990)]{Gerin1990} Gerin, M., Combes, F., \& Athanassoula, E. 1990, A\&A, 230, 37
\bibitem[{Geron et al.}(2025)]{Geron2025} G\'{e}ron, T., Smethurst, R. J., Dickinson, H., et al. 2025, ApJ, 987, 74
\bibitem[{Guo et al.}(2023)]{Guo2023} Guo, Y., Jogee, S., Finkelstein, S. L., et al. 2023, ApJ, 945, L10
\bibitem[{Guo et al.}(2025)]{Guo2025} Guo, Y., Jogee, S., Wise, E., et al. 2025, ApJ, 985, 181
\bibitem[{Hohl}(1971)]{Hohl1971} Hohl, F. 1971, ApJ, 168, 343
\bibitem[{Joshi et al.}(2020)]{Joshi2020} Joshi, G. D., Pillepich, A., Nelson, D., et al. 2020, MNRAS, 496, 2673
\bibitem[{Lansbury et al.}(2014)]{Lansbury2014} Lansbury, G. B., Lucey, J. R., \& Smith, R. J. 2014, MNRAS, 439, 1749
\bibitem[{Le Conte et al.}(2024)]{LeConte2024} Le Conte, Z. A., Gadotti, D. A., Ferreira, L., et al. 2024,
	MNRAS, 530, 1984
\bibitem[{{\L}okas}(2018)]{Lokas2018} {\L}okas, E. L. 2018, ApJ, 857, 6
\bibitem[{{\L}okas}(2020)]{Lokas2020} {\L}okas, E. L. 2020, A\&A, 638, A133
\bibitem[{{\L}okas}(2021)]{Lokas2021} {\L}okas, E. L. 2021, A\&A, 647, A143
\bibitem[{{\L}okas}(2024)]{Lokas2024} {\L}okas, E. L. 2024, A\&A, 687, A82
\bibitem[{{\L}okas}(2025a)]{Lokas2025a} {\L}okas, E. L. 2025a, A\&A, 700, A258
\bibitem[{{\L}okas}(2025b)]{Lokas2025b} {\L}okas, E. L. 2025b, A\&A, 702, A7
\bibitem[{{\L}okas}(2025c)]{Lokas2025c} {\L}okas, E. L. 2025c, ApJ, 991, L52
\bibitem[{{\L}okas et al.}(2014)]{Lokas2014} {\L}okas, E. L., Athanassoula, E., Debattista, V. P., et al. 2014,
        MNRAS, 445, 1339
\bibitem[{{\L}okas et al.}(2015)]{Lokas2015} {\L}okas, E. L., Semczuk, M., Gajda, G., \& D'Onghia, E. 2015,
	ApJ, 810, 100
\bibitem[{{\L}okas et al.}(2016)]{Lokas2016} {\L}okas, E. L., Ebrov\'{a}, I., del Pino, A., et al. 2016, ApJ, 826, 227
\bibitem[{Marinacci et al.}(2018)]{Marinacci2018} Marinacci, F., Vogelsberger, M., Pakmor, R., et al. 2018,
        MNRAS, 480, 5113
\bibitem[{Mastropietro et al.}(2005)]{Mastropietro2005} Mastropietro, C., Moore, B., Mayer, L., et al. 2005,
        MNRAS, 364, 607
\bibitem[{M\'{e}ndez-Abreu et al.}(2010)]{Mendez2010} M\'{e}ndez-Abreu, J., S\'{a}nchez-Janssen, R., \& Aguerri, J. A.
	L. 2010, ApJ, 711, L61
\bibitem[{M\'{e}ndez-Abreu et al.}(2023)]{Mendez2023} M\'{e}ndez-Abreu, Costantin, L., \& Kruk, S. 2023, A\&A, 678, A5
\bibitem[{Miwa \& Noguchi}(1998)]{Miwa1998} Miwa, T., \& Noguchi, M. 1998, ApJ, 499, 149
\bibitem[{Naiman et al.}(2018)]{Naiman2018} Naiman, J. P., Pillepich, A., Springel, V., et al., 2018, MNRAS, 477, 1206
\bibitem[{Nelson et al.}(2018)]{Nelson2018} Nelson, D., Pillepich, A., Springel, V., et al. 2018, MNRAS, 475, 624
\bibitem[{Nelson et al.}(2019a)]{Nelson2019a} Nelson, D., Springel, V., Pillepich, A., et al. 2019a,
        Computational Astrophysics and Cosmology, 6, 2
\bibitem[{Nelson et al.}(2019b)]{Nelson2019b} Nelson, D.,  Pillepich, A., Springel, V., et al. 2019b, MNRAS, 490, 3234
\bibitem[{Noguchi}(1987)]{Noguchi1987} Noguchi, M. 1987, MNRAS, 228, 635
\bibitem[{Ostriker \& Peebles}(1973)]{Ostriker1973} Ostriker, J. P., \& Peebles, P. J. E. 1973, ApJ, 186, 467
\bibitem[{Peschken \& {\L}okas}(2019)]{Peschken2019} Peschken, N., \& {\L}okas, E. L. 2019, MNRAS, 483, 2721
\bibitem[{Pillepich et al.}(2018)]{Pillepich2018} Pillepich, A., Nelson, D., Hernquist, L., et al. 2018,
        MNRAS, 475, 648
\bibitem[{Pillepich et al.}(2019)]{Pillepich2019} Pillepich, A., Nelson, D.,  Springel, V., et al. 2019,
        MNRAS, 490, 3196
\bibitem[{Rosas-Guevara et al.}(2020)]{Rosas2020} Rosas-Guevara, Y., Bonoli, S., Dotti, M., et al. 2020, MNRAS, 491,
        2547
\bibitem[{Rosas-Guevara et al.}(2022)]{Rosas2022} Rosas-Guevara, Y., Bonoli, S., Dotti, M., et al. 2022, MNRAS, 512,
	5339
\bibitem[{Rosas-Guevara et al.}(2025)]{Rosas2025} Rosas-Guevara, Y., Bonoli, S., Puchwein, E., Dotti, M., \&
	Contreras, S. 2025, A\&A, 698, A20
\bibitem[{Springel et al.}(2001)]{Springel2001} Springel, V., White, S. D. M., Tormen, G., \& Kauffmann, G. 2001,
	MNRAS, 328, 726
\bibitem[{Springel et al.}(2018)]{Springel2018} Springel, V., Pakmor, R., Pillepich, A., et al. 2018, MNRAS, 475, 676
\bibitem[{Tawfeek et al.}(2022)]{Tawfeek2022} Tawfeek, A. A., Cervantes Sodi, B., Fritz, J., et al. 2022, ApJ, 940, 1
\bibitem[{Thompson}(1981)]{Thompson1981} Thompson, L. A. 1981, ApJ, 244, L43
\bibitem[{Tremaine \& Weinberg}(1984)]{Tremaine1984} Tremaine, S., \& Weinberg, M. D. 1984, ApJ, 282, L5
\bibitem[{Zhao et al.}(2020)]{Zhao2020} Zhao, D., Du, M., Ho, L. C., Debattista, V. P., \& Shi, J. 2020, ApJ, 904, 170
\bibitem[{Zhou et al.}(2020)]{Zhou2020} Zhou, Z.-B., Zhu, W., Wang, Y., \& Feng, L.-L. 2020, ApJ, 895, 92


\end{thebibliography}
\end{document}